\documentclass[conference]{IEEEtran}
\usepackage[utf8]{inputenc}
\usepackage{graphicx}
\usepackage{xcolor}
\usepackage{url}
\usepackage{mdframed}
\usepackage{subfig}
\usepackage{tabularx}
\usepackage{multirow}
\usepackage{colortbl}
\usepackage{booktabs} % For formal tables

\begin{document}
\title{Requirements Intelligence with OpenReq Analytics}

\author{\IEEEauthorblockN{Christoph Stanik and Walid Maalej
\IEEEauthorblockA{University of Hamburg\\
Hamburg, Germany\\
\{stanik, maalej\}@informatik.uni-hamburg.de}}}

\maketitle

\begin{abstract}
With the rise of social media like Twitter and distribution platforms like app stores, users have various ways to express their opinions about software products.
Popular software vendors get user feedback thousandfold per day.
Research has shown that such feedback contains valuable information for software development teams.
However, a manual analysis of user feedback is cumbersome and hard to manage.
We present \emph{OpenReq Analytics}, a software requirements intelligence service, that collects, processes, analyzes, and visualizes user feedback.
\end{abstract}
\begin{IEEEkeywords} Requirements Intelligence, Data-Driven Requirements, Data Mining, Social Media Analytics, App Store Analytics \end{IEEEkeywords}
\IEEEpeerreviewmaketitle

%%%%%%%%%%%%%%%%%%%%%%%%%%%%%%%
% DOCUMENT START
%%%%%%%%%%%%%%%%%%%%%%%%%%%%%%%
\section{Introduction} \label{sec:introduction}
Software users share a large amount of feedback, which can be valuable to software development teams to better understand user needs.
Feedback shared in platforms such as app stores and Twitter contain insights like problems/bugs, feature requests, inquiries, or experience reports~\cite{Pagano:App:2013}.
Research showed that vendors considering user feedback are more successful in terms of download numbers and ratings, but a manual analysis is cumbersome \cite{palomba2015user}.
Further, interviews with industry practitioners highlighted the need for developers and managers for tool support to monitor user feedback continuously \cite{Maalej:REJ:2016}.
% In this work, we introduce \emph{OpenReq Analytics}, a tool for continuously collecting, processing, classifying, and visualizing user feedback.

%%%%%%%%%%%%%%%%%%%
% Section
%%%%%%%%%%%%%%%%%%%
\section{Requirements Intelligence} \label{sec:tool}
Inspired by Maalej et al. \cite{maalej2016toward}, we define (Software) Requirements Intelligence as a data-driven concept to explore requirements in the masses of implicit feedback (i.e., usage data) and explicit feedback (i.e., written feedback). 
As a step toward requirements intelligence, we are developing the web-based tool \emph{OpenReq Analytics}.
\emph{OpenReq}\footnote{\textbf{Website:} \url{www.openreq.eu}, \textbf{GitHub:} \url{https://github.com/OpenReqEU}} is a European Horizon 2020 open source project, which aims at creating methods, tools, and APIs toward achieving an intelligent recommendation and decision support for community-driven requirements engineering.
\emph{OpenReq Analytics} is the OpenReq component that collects, processes, analyzes, and visualizes implicit and explicit user feedback from app stores and Twitter. \emph{OpenReq Analytics} targets developers, as well as product and innovation managers. 
To get started, the users only need to configure the  accounts they want to analyze, e.g. in Twitter.

\subsection{Architecture} \label{ssec:architecture}
\begin{figure}[tb]
    \centering
    \includegraphics[width=\columnwidth]{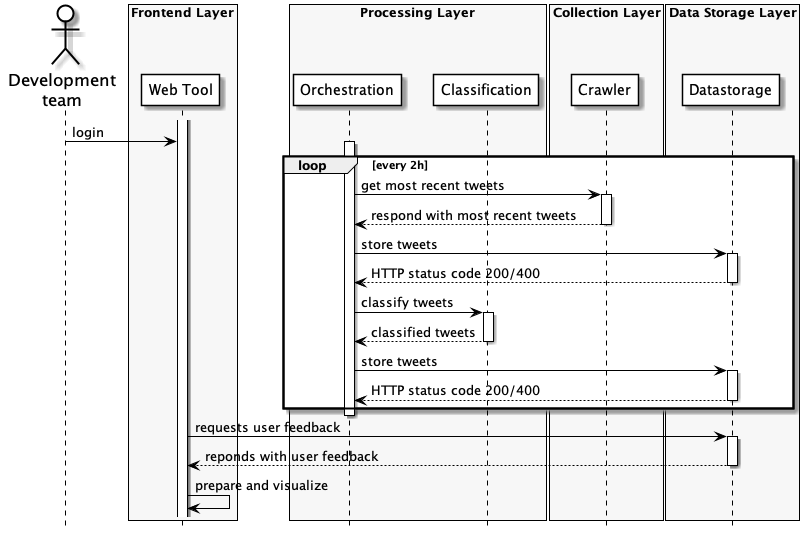}
    \caption{Microservice Architecture of \emph{OpenReq Analytics.}}
    \label{fig:architecture}
\end{figure}
\emph{OpenReq Analytics} is currently based on ten microservices.
Figure \ref{fig:architecture} shows the communication flow of the microservices for monitoring tweets from Twitter.
The single access point is the web-tool (see Fig. \ref{fig:dashboard}), which loads all data needed for visualization purposes from the \emph{Datastorage} microservice.
To ensure that the data presented is up-to-date, the orchestration microservices continuously crawl, classify, and store user feedback in a two-hour interval.

\subsection{Dashboard} \label{ssec:dashboard}
\begin{figure}[tb]
    \centering
    \includegraphics[width=\columnwidth]{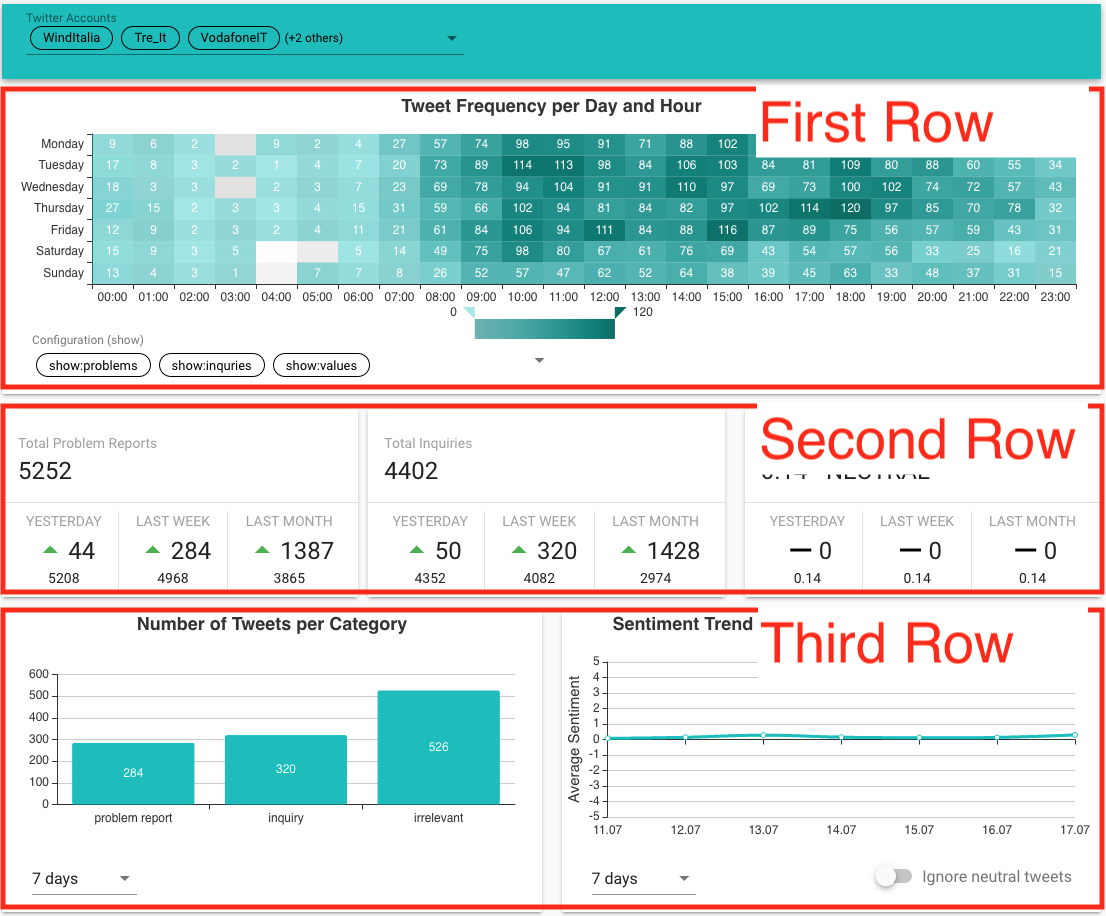}
    \caption{\emph{OpenReq Analytics}: The Dashboard.}
    \label{fig:dashboard}
\end{figure}
The main view of \emph{OpenReq Analytics} is the dashboard shown on Figure \ref{fig:dashboard}. It consists of three rows, each explained in the following paragraphs.

\textbf{When do users write?} 
The first row shows a heat map that gives insights on when users are giving feedback.
The heat map is an helpful indicator of when users face issues or think about new functionality. 
It assists vendors to understand when customer support is needed. 
When this explicit feedback analysis is combined with implicit feedback, detailed conclusions about usage patterns such as which steps caused the issue can be drawn.
\emph{OpenReq Analytics} allows filtering for \emph{relevant feedback entries}, which are only those that are either a problem report, feature request, or inquiry.

\textbf{Trend Reports.}
In the second row, three trend reports are showing information regarding predefined time windows.
The goal is to give an overview of three particular interests: 1) how many users are facing problems, 2) how many inquiries (incl. feature requests) are being received, and 3) what is the overall sentiment of the users when giving feedback?
This view helps to understand the general performance of the vendor over time.

\textbf{Historical Analysis.}
The third row enables the user to analyze custom time frames such as specific releases.
If, e.g., the software got a new release, the vendor can understand if bug reports are peaking or if less are being received.
On the other hand, whenever problem reports are peaking, users might get impatient and write rather negative reviews \cite{martens:sentiment:2017}. 
The sentiment chart on the right side helps to understand when the overall sentiment is back to normal, and if the strategy followed by the customer support is effective.

\subsection{Focus Views} \label{ssec:focus_views}
\begin{figure}[tb]
    \centering
    \includegraphics[width=\columnwidth]{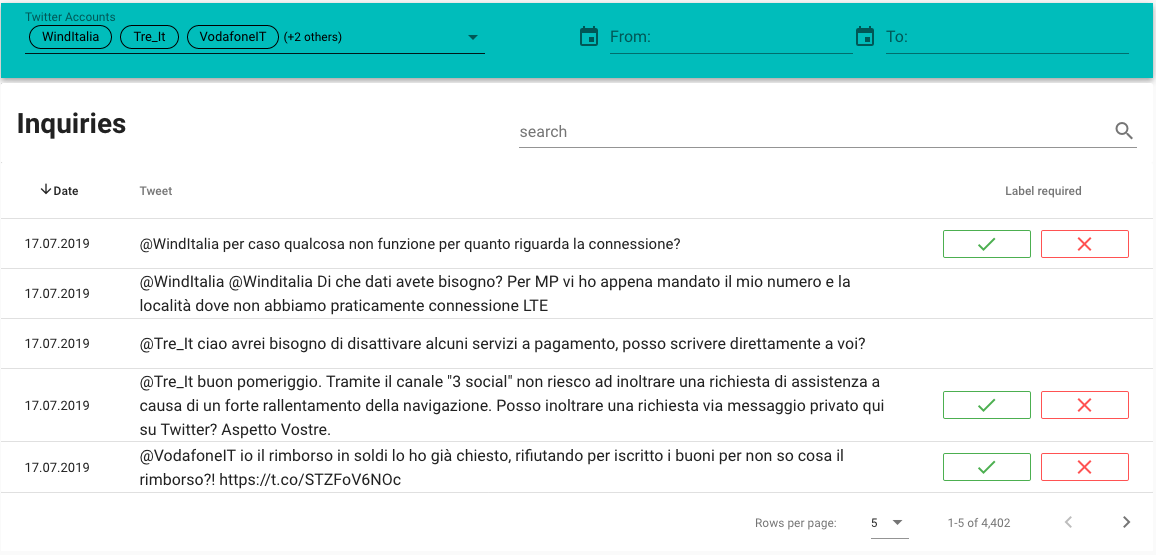}
    \caption{\emph{OpenReq Analytics}: Focus View for Inquiries.}
    \label{fig:focus_view}
\end{figure}
\emph{OpenReq Analytics} has two focus views (see Figure \ref{fig:focus_view})---each aggregating information related to either problem reports or inquiries.
The goals of the focus views are to aggregate feedback and to enable the development team to explore the concrete feedback given.
In this view, one can search for keywords of interest, filter the source of the feedback and its language, and defined time frames. 
A machine learning model performs the separation into problem reports and inquiries following best practices from related work \cite{Maalej:REJ:2016, Guzman:Twitter:2017}.
However, as the performance of machine learning models may decay over time,  e.g., because software introduced new features or the product portfolio of a vendor changed, it is important to keep the models up-to-date.
For this, we included an active learning component.
On the right side of Figure \ref{fig:focus_view}, two buttons are shown---one for agreeing with the classification and one for disagreeing.
When the software development team disagrees with the classification, \emph{OpenReq Analytics} shows other categories the user feedback should belong to.
In case the problem report focus view is open and the development team disagrees with the classification, two buttons appear: one for labeling that feedback as an inquiry and one for labeling it as irrelevant.
It is important to note, that only user feedback can be labeled for which the classification model is uncertain.
The rationale behind that decision is twofold.
First, those cases have a higher impact on the improvement of the model.
Second, software development teams are not overwhelmed with a labeling task of potentially thousands of labels per day.

%%%%%%%%%%%%%%%%%%%
% Section
%%%%%%%%%%%%%%%%%%%
\section{Conclusion and Future Work} \label{sec:conclusion}
\emph{OpenReq Analytics} is still under development and will get additional  features in the future.
One feature to come is topic modeling as at the moment the information presented is only separated in the high-level categories \emph{problem reports}, \emph{inquiries}, and \emph{irrelevant}.
But this leaves out details such as ``what particular feature causes the problem?'' or ``what concrete feature our user wish?'' \cite{Johann:2017:app}.

In addition to the historical analysis, we will include a release-based perspective which helps to understand better how each software/service update impacts user feedback.

Another import feature is clustering of similar user feedback to get a better understanding of how many users are facing a problem or requesting specific information or features.

%%%%%%%%%%%%%%%%%%%
% Section
%%%%%%%%%%%%%%%%%%%
\section*{Acknowledgement}
The work presented in this paper has been conducted within the scope of the Horizon 2020 project OpenReq, which is supported by the European Union under the Grant Nr. 732463

%%%%%%%%%%%%%%%%%%%%%%%%%%%%%%%
% DOCUMENT END
%%%%%%%%%%%%%%%%%%%%%%%%%%%%%%%
\bibliographystyle{abbrv}
\bibliography{lib}

\end{document}